\documentclass[fleqn,twoside]{article}
\usepackage{gc,amssymb,cite}

\newcommand{\be}[1]{\beq\label{#1}}
\newcommand{\ber}[1]{\bear\label{#1}}
\newcommand{\berr}[1]{\bearr\label{#1}}

\newcommand{\R}{ \mathbb{R} }

\heads
   {H. Dehnen, V.D. Ivashchuk,
   S.A. Kononogov and V.N. Melnikov}
  {On time variation of $G$ in multidimensional models with two curvatures}

\bls{0.95}
\begin{document}
\twocolumn[
\prepno{gr-qc/0602108}{\GC{11} 340 (2005)}

\Title
  {On time variation of $G$ in multidimensional models\yy
  with two curvatures}

\Aunames{
   H. Dehnen\auth{1,a}, V.D. Ivashchuk\auth{2,b,c},
   S.A. Kononogov\auth{3,b} and V.N. Melnikov\auth{4,b,c}
   }

\Addresses{
 \addr a {Universit\"at Konstanz, Fakult\"at f\"ur
     Physik, Fach  M 568, D-78457, Konstanz}
 \addr b {Centre for Gravitation and Fundamental Metrology,
     VNIIMS, 46 Ozyornaya St., Moscow 119361, Russia}
 \addr c {Institute of Gravitation and Cosmology,
     Peoples' Friendship University of Russia,
 	6 Miklukho-Maklaya St., Moscow 117198, Russia}
	}


\Abstract
 {Expressions for $\dot G$ are considered in a multidimensional model with
 an Einstein internal space and a multicomponent perfect fluid. In the case
 of two non-zero curvatures without matter, a mechanism for prediction of
 small $\dot G$ is suggested. The result is compared with exact
 (1+3+6)-dimensional solutions. A two-component example with two matter
 sources (dust + 5-brane) and two Ricci-flat factor spaces is also
 considered.  }


] 
\email 1 {Heinz.Dehnen@uni-konstanz.de}
\email 2 {rusgs@phys.msu.ru}
\email 3 {kononogov@vniims.ru}
\email 4 {melnikov@phys.msu.ru}

 \section{Introduction}

 The idea of possible slow (cosmological) time variations of fundamental
 physical constants, the gravitational constant $G$ in particular, came out
 from Dirac's analysis in 1937 of some relations between macro- and
 micro-world phenomena.  His Large Numbers Hypothesis (LNH) was the origin
 of many further theoretical and experimental explorations of time-varying
 $G$. According to the LNH, $\dot{G}/G$ should have approximately the Hubble
 rate. Although it has become clear in the recent decades that the Hubble
 rate is too high to be compatible with experiment, the enduring legacy of
 Dirac's bold stroke is the acceptance by modern theories of non-zero values
 of $\dot{G}/G$ as being potentially consistent with physical reality.

 After Dirac's original hypothesis, some new ideas appeared as well as
 generalized theories of gravity admitting variations of the effective
 gravitational coupling.

 Different theoretical schemes lead to temporal variations of the
 effective gravitational constant:

 \begin{enumerate}
 \item
 Empirical models and theories of Dirac type, where $G$ is replaced with
 $G(t)$.

 \item
 Numerous scalar-tensor theories of Jordan-Brans-Dicke type where $G$
 depends on the scalar field $\sigma (t)$.

 \item
 Gravitational theories with a conformal scalar field arising in different
 approaches \cite{mel76,Stan}.

 \item
 Multidimensional unified theories arising from supergravities and
 superstrings and a future possible M-theory, containing are dilaton fields
 and effective scalar fields appearing in our 4-dimensional spacetime from
 extra dimensions \cite{Mel}. They may also help in solving the problem of a
 variable cosmological constant from Planckian to present values.
\end{enumerate}

 A striking feature of most modern scalar-tensor and  unification theories
 is that they do not admit a unique and universal constant values of
 physical constants, including the Newtonian gravitational coupling
 constant $G$. In this paper, we briefly set out the results of some
 calculations which have been carried out for various theories, and discuss
 various bounds that may be suggested by multidimensional theories.
 Although the bounds on $\dot G$ and $G(r)$ are, in some classes of theories,
 rather wide on purely theoretical grounds as a result of adjustable
 parameters, we note that the observational data concerning other phenomena
 may place limits on the possible range of these adjustable parameters.

 Here we restrict ourselves to the problem of $\dot G$ (for $G(r)$ see
 \cite{mel76,Stan,mel-Erice,4,5,Mel}). We show that various theories predict
 $\dot{G}/G$ of the order $10^{-12}$/yr or smaller. The significance of
 this fact for experimental and observational determinations of the value of
 or upper bound on $\dot G$ is the following:  any determination with error
 bounds significantly below $10^{-12}$ will be typically compatible with
 only a small portion of the existing theoretical models and will therefore
 cast serious doubt on the viability of all other models. In short, a tight
 bound on $\dot G$, in conjunction with other astrophysical observations,
 will be a very effective ``theory killer''. In other words, it may be
 {\it a new test of cosmological models in addition to standard
 cosmological tests.}

 Some estimations for $\dot G$ were made long ago in the framework of GR
 with a conformal scalar field \cite{BMS-ITP,ZM-79} and in general
 scalar tensor theories using the values of the cosmological parameters
 ($\Omega$, $H$, $q$ etc.) \cite{Stan,BMN}. With modern values, they predict
 $\dot{G}/G$ at the level of $10^{-12}$/yr and smaller (see also recent
 estimations by A. Miyazaki \cite{Mi}, predicting time variations of $G$ at
 the level of $10^{-13}$/yr) for a Machian-type cosmological solution in the
 Brans-Dicke theory and Fujii's estimations \cite{Fj}.

 The most reliable experimental bounds on $\dot{G}/G$ (radar ranging of
 spacecraft dynamics \cite{Hel,P-97,N-03}) and laser lunar ranging
 \cite{Dic} give the limit of $10^{-12}/yr$). There also exist some
 model-dependent measurements of $\dot G$ from Big Bang nucleosynthesis at
 the level of a few units of $10^{-13}/{\rm yr}$.

\section{$\dot G$ in (1+3+$N$)-dimensional cosmology
          with a multicomponent anisotropic fluid }

 We consider here a $(4+N)$-dimensional cosmology with an isotropic 3-space
 and an Einstein internal space. The Einstein equations provide a relation
 between ${\dot{G}/G}$ and other cosmological parameters.

\subsection{The model}

 Let us consider a $(4+N)$-dimensional theory with the gravitational action
\beq
    S_g = \frac{1}{2\kappa^2}\int d^{4+N}x\sqrt{-g}R,
\eeq
 where $\kappa^2$ is the fundamental gravitational constant. Then the
 gravitational field equations are
\beq
  R^M_P = \kappa^2\biggl(T^M_P-\delta^M_P\frac{T}{N+2}\biggr) ,
\eeq
 where $T^M_P$ is the $(4+N)$-dimensional energy-momentum tensor,
 $T=T^M_M$ and $M,P=0,...,N+3$.

 For the $(4+N)$-dimensional manifold we assume the structure
\beq
      M^{4+N} = \R_{*} \times M^3_k\times K^N\,
\eeq
 where $\R_{*}$ is the 1-dimensional time manifold, $M^3_k$ is a 3D space of
 constant curvature, $M^3_k = S^3,\ R^3,\ L^3$ for $k=+1,0,-1$,
 respectively, and $K^N$ is an $N$-dimensional Einstein manifold.

The metric is taken in the form
\bearr
	ds^2 =  g_{MN}dx^Mdx^N
\nnn \nq
	= - dt^2 + a^2(t)g^{(3)}_{ij}(x)dx^idx^j +
  		b^2(t)g^{(N)}_{mn}(y)dy^m dy^n ,
\ear
where $i,j,k = 1,2,3$;\ $m,n,p = 4,\ ...,\ N+3$;\ $g^{(3)}_{ij}$,\
$g^{(N)}_{mn}$,\ $a(t)$ and $b(t)$ are the metrics and scale factors
of $M^3_k$ and $K^N$, respectively.

For $T^M_P$ we adopt the expression for a multicomponent (anisotropic) fluid
\beq
  (T^M_P) = \sum_{\alpha = 1}^m
  {\rm diag}(- \rho^{\alpha} (t),  p_3^{\alpha}(t)\delta^i_j,
       p_N^{\alpha}(t)\delta^m_n).
\eeq

Under these assumptions, the Einstein equations take the form
\berr{6}
 \frac{3\ddot{a}}{a} + \frac{N \ddot{b}}{b} =
 		\frac{\kappa^2}{N+2} \sum_{\alpha = 1}^m
 				[-(N{+}1)\rho^{\alpha}
	 		\!- 3p_3^{\alpha} -N p_N^{\alpha}],
\\ \lal  \label{7}
 	\frac{\ddot{a}}{a} + \frac{2 \dot{a}^2}{a^2} +
 		\frac{N \dot{a} \dot{b}}{ab}   +  \frac{2k}{a^2}
\nnn \cm
 	= \frac{\kappa^2}{N+2}  \sum_{\alpha = 1}^m [\rho^{\alpha} +
 			(N-1) p_3^{\alpha} - N p_N^{\alpha}],
\\ \lal   \label{8}
 	\frac{\ddot{b}}{b} + (N-1) \frac{\dot{b}^2}{b^2} +
 		\frac{3 \dot{a} \dot{b}}{ab} +  \frac{\lambda}{b^2}
\nnn \cm
	=  \frac{\kappa^2}{N+2}  \sum_{\alpha = 1}^m
 		[\rho^{\alpha} -3 p_3^{\alpha}+ 2 p_N^{\alpha}].
\ear
Here
\beq       \label{8a}
  	R_{m n}[g^{(N)}] = \lambda g^{(N)}_{mn},
\eeq
 $m,n = 1,\ldots, N$, where $\lambda$ is constant.

The 4-dimensional density is
\be{9}   \nq
 \rho^{\alpha,(4)}(t) = \int_K d^Ny\sqrt{g^{(N)}}b^N(t)\rho^{\alpha}(t) =
 		\rho^{\alpha}(t) b(t),
\eeq
 where we have normalized the factor $b(t)$ by putting
\be{10}
	   \int_K d^Ny\sqrt{g^{(N)}} = 1.
\eeq

 On the other hand, to get the 4D gravity equations, one should put $8\pi
 G(t)\rho^{\alpha(4)}(t) = \kappa^2\rho^{\alpha}(t)$.  Consequently, the
 effective 4D gravitational ``constant'' $G(t)$ is defined by
\beq
   	8\pi G(t) = \kappa^2b^{-N}(t)
\eeq
 whence its time variation is expressed as
\be{G-dot}
 		 \dot{G}/G = -N \dot{b}/b.
\eeq

 \subsection{Cosmological parameters}

 Some inferences concerning the observational cosmological parameters can
 be extracted directly from the equations without solving them \cite{BIM1}.
 Indeed, let us define the Hubble parameter $H$, the density parameters
 $\Omega^{\alpha}$ and the ``deceleration'' parameter $q$ referring to a
 fixed instant $t_0$ in the usual way:
\beq \label{cosm}\nq
	H = \frac{\dot a} a,  \quad\
	\Omega^{\alpha} = \frac{8\pi G\rho^{\alpha,(4)}}{3H^2} =
 	\frac{\kappa^2\rho^{\alpha}}{3H^2}, \quad\
	q = -\frac{a\ddot a}{{\dot a}{}^2}.
\eeq
 Besides, instead of $G$, let us introduce the dimensionless parameter
\be{g-dot}
	  g = \dot{G}/GH = -Na \dot{b}/\dot{a}b .
\eeq

The present observational upper bound on $g$ is
\be{18a}
   		g  < 0.1
\eeq
 if we take in accord with  \cite{Hel,Dic}
\be{18b}
	  \dot{G}/G  <   0.6 \times 10^{-11}({\rm yr}^{-1})
\eeq
 and $H = (0.7 \pm 0.1) \times 10^{-11}/{\rm yr}
 	  \approx 70 \pm 10$ km/(s. Mpc).

\section{A vacuum model with two Einstein spaces}

 Here we consider the vacuum case when $T^M_P = 0$.
 Let us suppose that $t_0$ is an extremum point of the function $b(t)$, i.e.,
\be{2.1}
    \dot{b}(t_0) = 0.
\eeq
 At this point we get $\dot{G}(t_0) = 0$.
 From \eqs (\ref{6}), (\ref{7}), (\ref{8}) we obtain that for $t =t_0$
\berr{2.6}
	  \frac{3\ddot{a}}{a} + \frac{N\ddot{b}}{b} = 0,
\\    \lal   \label{2.7}
	  \frac{\ddot{a}}{a} = -\frac{2 \dot{a}^2}{a^2} - \frac{2k}{a^2} =0,
\\    \lal                            \label{2.8}
 	\frac{\ddot{b}}{b} = -  \frac{\lambda}{b^2}.
\ear

 Suppose that ``we live'' near the point $t_0$, then, according to modern
 observations on the acceleration of the Universe expansion
 (\cite{Riess,Perl}) we should put
\berr{2.10a}
     		\dot{a}(t_0) > 0
\\ \lal  \label{2.10}
     		\ddot{a}(t_0) > 0.
\ear
 This implies
\be{2.11}
	     k < 0
\eeq
  due to (\ref{2.7}) and
\be{2.12}
	   \ddot{b}(t_0) < 0,  \qquad  \lambda > 0
\eeq
 due to  (\ref{2.6}) and (\ref{2.8}). Thus our 3-dimensional space should
 have a negative curvature while the internal $N$-dimensional space should
 have a positive curvature.

 From  (\ref{2.6})-(\ref{2.8}) we obtain,
 using the definitions of the cosmological parameters,
\ber{2.13}
  	\frac{|2k|}{H_0^2 a_0^2} \eql  2 + |q_0|,
\\  \label{2.14}
  	\frac{d_2| \lambda|}{H_0^2 b_0^2} \eql  3 |q_0|.
\ear
 Here $a_0 = a(t_0)$  and  $b_0 = b(t_0)$. Since by assumption ``we live''
 near the point $t_0$, we get
\be{2.15}
  	\frac{\dot{b}}{b}  \approx \frac{\ddot{b}_0 (t - t_0)}{b_0},
\eeq
  and due to (\ref{G-dot}) and (\ref{2.6}) we find
\be{2.16}
  \frac{\dot{G}}{G} = - N \dot{b}/b  \approx
  - N \frac{\ddot{b}_0}{b_0} (t - t_0) = 3 \frac{\ddot{a}_0 (t - t_0)}{a_0}.
\eeq
 The subscript "0" refers to $t_0$. Using the definitions of the
 cosmological parameters (\ref{cosm}), we obtain in our approximation
 \be{2.17}
  	\frac{\dot{G}}{G} \approx  - 3 q_0 H_0^2 (t - t_0).
 \eeq
  Recall that $q_0 < 0$, hence $\dot{G}/G > 0$ for $t > t_0$ and
  $\dot{G}/G < 0$ for $t < t_0$.

  We also note that, in our approximation,$\dot{G}/G$ is independent of
  the internal space dimension $N = \dim K$.

\subsection{Exact $1 + 3 + 6$ solution}

  Now we consider an exact solution from Ref. \cite{GIM}
  defined on the manifold
\beq  \label{2.18}
  		M = \R_{*} \times M^{(3)} \times M^{(6)},
\eeq
 with the metric
\bearr   \label{2.19}
      ds^2 =  \left(f_1 f_2 \right)^{- \frac{1}{2}}
        [ - 2 f_1^{-2}  (d\tau)^2
\nnn
	+ |\lambda_3| g^{(3)}_{ij}(x)dx^idx^j
         	+ f_2 |\lambda_6| g^{(N)}_{mn}(y)dy^mdy^n],
\ear
 where $(M^{(3)},\,g^{(3)})$ and $(M^{(6)},\,g^{(6)})$ are Einstein spaces:
\beq                                                   \label{2.20}
        {\rm Ric}\, [g^{(i)}]  = \lambda_i g^{(i)},
\eeq
 $i = 3, 6$. We use the notations $\lambda_3 = 2k$ and $\lambda_6 = \lambda$.
 In (\ref{2.19}),
\bear            \label{2.21}
  f_1 \eql \left |\tau^2 + \eps_3 \right|,
\\     \label{2.22}
  f_2 \eql - 3 \eps_6 (\tau^2 {+} \eps_3)  \{ 1
    	+ \tau [h(\tau, \eps_3)+ C_1] \} + \eps_3 \eps_6 > 0,
\nnn
\ear
 where $C_1 = \const$, $\eps_i = \sign (\lambda_i)$, $i = 3, 6$, and
\bear                                                         \label{2.23}
    h(\tau, \eps_3) \eql \frac{1}{2} \ln\left|\frac{\tau-1}{\tau+1}\right|, \
           \qquad \eps_3 = - 1,
\\     \label{2.24}
   h(\tau, \eps_3) \eql  \arctan(\tau), \ \qquad \eps_3 = 1.
\ear

  As was mentioned above, we should restrict our consideration to the case
  when ``our'' 3-dimensional space has a negative curvature while the
  6-dimensional ``internal'' space has a positive curvature, i.e.,
\beq                                                        \label{2.25}
     	\eps_3 = - 1, \qquad 	\eps_6 =  1.
\eeq

  The analysis carried out in \cite{GIM} shows that the scale factor of
  	``our'' 3-space
  \beq                          \label{2.26}
     	a_3 = a = (f_1 f_2)^{-1/4}|\lambda_3|^{1/2}
  \eeq
    has a minimum at some point $\tau_{*}$ when the branch of the solution
    with $\tau \in (\tau_{-}, \tau_{+})$ is considered. Here $\tau_{-}$ and
    $\tau_{+}$ are roots of the equation $f_2(\tau) = 0$ belonging to the
    interval $(0,\ 1)$. In this case, the scale factor of ``our'' space
    $a_3(\tau)$ monotonically decreases in the interval $(\tau_{-}, \tau_{*})$
    and monotonically increases in the interval $(\tau_{*}, \tau_{+})$.

    The scale factor  of the ``internal'' 6-space
   \beq                                      \label{2.27}
     	a_6 = b = (f_1 f_2)^{-1/4} f_2^{1/2} |\lambda_6|^{1/2}
   \eeq
    has a maximum at some point $\tau_{0}$. It monotonically increases in
    the interval $(\tau_{-}, \tau_{0})$ and monotonically decreases in the
    interval $(\tau_{0}, \tau_{+})$.

\medskip\noi
   {\bf Remark}. For other branches of the solution with either $\tau \in
   (\tau_{-}, \tau_{+ \infty})$ or  $\tau \in (- \infty, \tau_{+}),
   (|\tau_{-}|, |\tau_{+}| > 1)$ we get a monotonic behaviour of both scale
   factors $a_3(\tau)$ and $a_6(\tau)$.

   Now consider our solution in synchronous time:
\bearr                                                         \label{2.28}
   	ds^2 = - dt^2 + a_3^2(t) g^{(3)}_{ij}(x)dx^idx^j
\nnn \inch
        	 + a_6^2(t) g^{(N)}_{mn}(y)dy^mdy^n,
\ear
 where
\beq
  t_s=  \sqrt{2} \int ^{\tau}_{\tau_{-}}d\tau'(f_1f_2)^{-1/4} f_1^{-1} .
\eeq
  The function $t_s(\tau)$ is monotonically increasing from $t_s(\tau_{-}) =
  0$ to $T = t_s(\tau_{+})$.

  The 3-space scale factor has a minimum at the point $t_{0} = t(\tau_{*})$.
  The function $a_3(t)$ monotonically decreases from infinity to finite value
  in the interval $(0, t_{0})$ and monotonically increases to infinity
  in the interval $(t_{0}, T)$.

  The 6-space scale factor has a maximum at the point $t_{*} = t(\tau_{*})$.
  The function $a_6(t)$ monotonically increases from zero to a finite
  value in the interval $(0, t_{*})$ and monotonically decreases
  to zero in the interval $(t_{*}, T)$.

  Only in case $C_1 > 0$ we get $t_{*} < t_0$ and hence in the ``epoch''
  near $t_0$ we get an accelerating expansion of ``our'' 3-space.

\section{A model with two Ricci-flat spaces and a two-component fluid}

 Here we consider another example when two factor spaces are Ricci-flat.
 In this case, excluding $b$ from (\ref{6}) and (\ref{8}), we get
\be{14}
 	\frac{N-1}{3N} g^2 -g +
  		q - \sum_{\alpha = 1}^m A^{\alpha} \Omega^{\alpha} = 0
\eeq
with
\be{15}
  A^{\alpha} = \frac{1}{N+2} [2N+1+3(1-N)\nu_3^{\alpha} + 3N \nu_N^{\alpha}],
\eeq
where
\be{16}
 	\nu_3^{\alpha} = p_3^{\alpha}/\rho^{\alpha}, \qquad
    \nu_N^{\alpha} = p_N^{\alpha}/\rho^{\alpha}, \qquad   \rho^{\alpha} > 0.
\eeq
 When $g$ is small, we get from (\ref{14})
 \be{17}
   	g \approx q  -  \sum_{\alpha = 1}^m A^{\alpha} \Omega^{\alpha}.
 \eeq

Note that (\ref{17}) for $N=6$, $m=1$, $\nu_3^{1}=\nu_6^{1}=0$ (so that
$A^{1}=13/8$) coincides with the corresponding Wu and Wang's relation
\cite{WW} obtained for large times in case $k=-1$ (see also \cite{IM1}).

If $k=0$, then in addition to (\ref{17}), one can obtain a separate relation
between $g$ and $\Omega^{\alpha}$, namely,
\be{18}
    \frac{N-1}{6N} g^2 - g + 1 - \sum_{\alpha = 1}^m \Omega^{\alpha} = 0
\eeq
(this follows from the Einstein equation $R^0_0-\frac{1}{2}R = \kappa^2
T^0_0$, which is certainly a linear combination of (\ref{6})-(\ref{8}).

\subsection{A two-component example: dust + $(N-1)$-brane}

 Let us consider a two component case: $m= 2$ \cite{IM-Was}. Let
 the first component (called ``matter'') be dust, i.e.
\be{19}
  		\nu_3^{1}  = \nu_N^{1} = 0,
\eeq
while the second one (called ``quintessence'') be an $(N-1)$-brane, i.e.,
\be{20}
  	\nu_3^{2}  =  1, \qquad  \nu_N^{2} = - 1.
\eeq

 We remind the reader that, as was mentioned in \cite{IMbil}, a
 multidimensional cosmological model on product manifold $\R \times M_1
 \times ... \times M_n$ with fields of forms (for a review see \cite{IMtop})
 may be described in terms of a multicomponent ``perfect'' fluid \cite{IMfl}
 with the following equations of state for an $\alpha$-s component:
 $p_i^{\alpha} =  -\rho^{\alpha}$ if the $p$-brane world volume contains
 $M_i$ and $p_i^{\alpha} = \rho^{\alpha}$ otherwise. Thus the field of form
 matter leads either to a $\Lambda$-term or to stiff matter equations of
 state in the internal spaces.

 In this case we get from (\ref{17}) for small $g$
\be{21}
  g \approx q  -\frac{2N+ 1}{N+2} \Omega^{1}+ 4 \frac{N - 1}{N+2} \Omega^{2},
\eeq
and for $k =0$ and small $g$ we obtain from (\ref{18})
\beq
  	1 -g  \approx \Omega^{1} + \Omega^{2}.
\eeq

Now we illustrate the formulas by the following example
when  $N =6$ ($K^6$ may be a Calabi-Yau manifold) and
\be{22}
  	-q  =  \Omega^{1} = \Omega^{2} = 0.5.
\eeq
We get from (\ref{21})
\be{23}
  	g \approx - \frac{1}{16} \approx -0.06
\eeq
in agreement with (\ref{18a}). In this case the second fluid component
corresponds to a magnetic (Euclidean) $NS5$-brane (in $D=10$ type I, Het or
II A string models). We here consider for simplicity the case of a constant
dilaton field.

This example tells us that, for a small enough temporal variation of $G$, we
may find estimates for $\dot G$ without considering of exact solutions.
But we should select the solutions that give us an accelerated expansion of
our world. We may use, for instance, the mechanism suggested above (see
\sect 3), but, instead of two curvatures, we should consider two fluid
components. This may be a subject of a separate study.

\section{Conclusions}

In this paper we have considered a multidimensional cosmological model with
an $m$-component anisotropic (``perfect'') fluid.  The multidimensional
Hilbert-Einstein equations led to relations between $\dot G$ and
cosmological parameters.

In the case of two non-zero curvatures without matter, we have suggested a
mechanism for predicting small $\dot G$. We conjectured that we ``live''
near the point $t_0$ where the time variation of $G$ is zero. When the
3-space has a negative curvature and the internal space has a positive
curvature, we get, in the vicinity of $t_0$, an accelerating expansion of
``our'' 3-dimensional space and a small value of $\dot{G}/G$ (see
(\ref{2.17})). We have shown that this result is compatible with the exact
$1+ 3 + 6$ solution  from \cite{GIM}. Recall that there only three exact
solutions are known for a vacuum cosmological model with a product of two
Einstein spaces, see \cite{GIM}. Recently, a certain interest in models
of this type has appeared in the  context of the acceleration problem, see
\cite{CHNOW} and references therein, but without any usage of exact solutions.

We have presented another example where two factor spaces are Ricci-flat
and, for a  wo-component example (dust + 5-brane) we have obtained a small
enough variation of $G$. This example may be used in a future work
for a generalization of the mechanism suggested for the vacuum case to
models with matter sources.

\Acknow
 The work of V.D.I. and V.N.M. was supported in part by DFG grant
 436RUS113/807/0-1 and by the Russian Foundation for Basic Researchs grant
 No. 05-02-17478. V.N.M. thanks the colleagues from the Physical Department
 of the University of Konstanz for hospitality during his visit in
 October-November 2005.

\end{document}